# A Multidimensional Hierarchical Framework for Modeling Speed and Ability in Computer-based Multidimensional Tests

Peida Zhan (Zhejiang Normal University)[1]

Hong Jiao (University of Maryland, College Park)

Wen-Chung Wang (The Education University of Hong Kong)

Kaiwen Man (University of Maryland, College Park)

**Note:**

[1] Corresponding author: Peida Zhan, Department of Psychology, College of Teacher Education, Zhejiang Normal University, No. 688 Yingbin Road, Jinhua, Zhejiang, 321004, P. R. China. Email: pdzhan@gmail.com

## Abstract

In psychological and educational computer-based multidimensional tests, latent speed, a rate of the amount of labor performed on the items with respect to time, may also be multidimensional. To capture the multidimensionality of latent speed, this study firstly proposed a multidimensional log-normal response time (RT) model to consider the potential multidimensional latent speed. Further, to simultaneously take into account the response accuracy (RA) and RTs in multidimensional tests, a multidimensional hierarchical modeling framework was proposed. The framework is an extension of the van der Linden (2007; doi:10.1007/s11336-006-1478-z) and allows a "plug-and-play approach" with alternative choices of multidimensional models for RA and RT. The model parameters within the framework were estimated using the Bayesian Markov chain Monte Carlo method. The 2012 Program for International Student Assessment computer-based mathematics data were analyzed first to illustrate the implications and applications of the proposed models. The results indicated that it is appropriate to simultaneously consider the multidimensionality of latent speed and latent ability for multidimensional tests. A brief simulation study was conducted to evaluate the parameter recovery of the proposed model and the consequences of ignoring the multidimensionality of latent speed.

With the popularity of computer-based tests, collection of item response times (RTs) has become a routine activity in large- and small-scale tests. For example, the Program for International Student Assessment (PISA) started using computer-based tests and recorded RTs data since the year of 2012. In addition to response accuracy (RA), RTs provide an additional source of information about working speed of respondents and time cost of items (Marianti, Fox, Avetisyan, Veldkamp, & Tijmstra, 2014; van der Linden, 2006, 2007, 2009; Zhan, Jiao, & Liao, 2018; Zhan, Liao, & Bian, 2018). Before making inferences by employing RTs, it is necessary to create an appropriate statistical model for RTs. Over the past few decades, various RT models have been presented based on cognitive/psychological theories and experimental research (for review, see Lee & Chen, 2011; Schnipke & Scrams, 2002; van der Linden, 2009). Conventionally, the speed-accuracy trade-off (Luce, 1986) was the main motivation for RT modeling, such as Thissen (1983), Wang and Hanson (2005), and Ferrando and Lorenzo-Seva (2007). However, the trade-off reflects only a within-person level relationship between speed and accuracy (van der Linden, 2009), which is hard to be evaluated based on single time-point (or cross-sectional) assessments (see Curran & Bauer, 2010; Molenaar, P. C., 2004). Typically, for a fixed set of items, once a respondent's working speed is fixed, his/her accuracy remains constant, therefore, the speed and accuracy are suggested to be modeled separately and their relationship can be modeled at a higher-level (van der Linden, 2006; 2007; 2009). To this end, van der Linden (2007) proposed a Bayesian hierarchical modeling framework, which is one of the most flexible tools to explain the relationship between response speed and accuracy. When comparing various RT models, Suh (2010) claimed that the Bayesian hierarchical modeling framework presents the most reasonable outcomes in both real and simulated data. In addition, some subsequent studies followed the thought of Bayesian hierarchical modeling framework, but treated the item effects as fixed (e.g., Molenaar, D., Tuerlinckx, & van der Maas, 2015; Wang, Chang, & Douglas, 2013; Wang & Xu, 2015). For simplicity, they are collectively referred to as *hierarchical modeling*. The hierarchical modeling is gaining more and more recognition and it is sufficiently generalized to integrate available measurement models for response accuracy and RTs (Fox & Marianti, 2016; Klein Entink, Fox, & van der Linden, 2009; Klein Entink, van der Linden, & Fox, 2009; Wang, et al., 2013; Wang & Xu, 2015; Zhan, Jiao et al., 2018; Zhan, Liao et al., 2018).

Currently, however, based on the hierarchical modeling, most researches only focus on unidimensional tests (e.g., van der Linden, 2007; Klein Entink, Fox, van der Linden, 2009; Klein Entink, van der Linden, et al., 2009; Meng, Tao, & Chang, 2015; Wang & Xu, 2015; Molenaar, D., Oberski, Vermunt, & De Boeck, 2016; Fox, Klein Entink, & Timmers, 2014). And only a unidimensional latent ability and a unidimensional latent speed are taken into account by using unidimensional item response theory (UIRT) models and unidimensional RT (URT) models, respectively, as shown in Figure 1(a).

In reality, respondents are likely to bring multiple latent abilities to bear when responding to items; meanwhile, items are likely to require various latent abilities to determine a correct response, especially in multidimensional tests (e.g., Reckase, 2009; Whitely, 1980; Tatsuoka, 1983). In addition, with the increasing demand for providing more detailed and refined feedback to test-takers, multidimensional tests have received much attention from practitioners and researchers.

In psychological and educational measurements, an appropriate notion of latent speed on test items is that of speed of labor (van der Linden, 2009). Therefore, latent speed can be defined as a rate of the amount of labor performed on the items with respect to time (van der Linden, 2011). Actually, the definition of latent speed should be discussed in a certain dimension of latent ability, because using the required latent ability is the basis of an effective labor. Due to the multidimensional nature of latent abilities and different items may require different kinds of abilities, latent speed may also be a multidimensional concept, each dimension of which corresponds to a specific type of labor or latent ability. For example, the latent speed corresponding to the latent ability of decoding (or algebra problem solving) may be different from the latent speed corresponding to the latent ability of encoding (or geometry problem solving). For another example, when respondents, especially for non-native English speakers, take part in the GRE® Subject Test (e.g., Mathematics), at least two abilities are needed, one for understanding the questions (e.g., the English reading ability), and one for solving the questions (e.g., the subject ability). Meanwhile, corresponding two latent speeds worked, one reflects the working speed of reading, and one reflects the working speed of problem-solving or applying subject ability.

Currently, although multidimensional models for response accuracy have been well developed (see Reckase, 2009), to our knowledge, there is a lack of multidimensional models for RTs to take account of the potential multidimensionality of latent speed. Recently, based on the hierarchical modeling, a few studies have attempted to use multidimensional models for response accuracy to capture the multidimensional structure of latent ability when multidimensional tests are involved; but only a URT model is used to capture the potentially multidimensional latent speed, as shown in Figure 1(b). For instance, Man, Jiao, Zhan, and Huang (2017) employed a compensatory multidimensional IRT model for response accuracy and the unidimensional log-normal RT model (van der Linden, 2006) for RTs. In addition, Zhan, Jiao et al. (2018) proposed a joint cognitive diagnosis modeling to simultaneously analysis RA and RTs in cognitive diagnosis. This approach was further extended to account for the paired local item dependence (Zhan, Liao et al., 2018). However, in these studies, because of the lack of multidimensional RT (MRT) models, only the relationship among multiple latent abilities and one single latent speed can be evaluated. Logically, as aforementioned, different latent abilities may be associated with different latent speeds. Thus, assuming the latent speeds corresponding to different latent abilities to be identical, as done by Man et al. (2017) and Zhan, Jiao et al. (2018), may be too restrictive to describe complicated data and thus may lead to biased conclusions. It is desirable to release this limitation to allow each latent ability to be associated with its own latent speed. As URT models may be inappropriate in practice, it is critical to develop a new RT model that considers multidimensionality of latent speed.

To meet the demand, we firstly extend the most popular unidimensional log-normal RT model (ULRTM; van der Linden, 2006) to be multidimensional and call it the *multidimensional log-normal RT model* (MLRTM). Secondly, a *multidimensional hierarchical modeling framework* for modeling multidimensional latent speed and multidimensional latent ability was proposed, as shown in Figure 1(c). The rest of the paper starts with a brief review of the ULRTM, followed by the presentation of the proposed MLRTM. Further, the proposed modeling framework and a corresponding joint multidimensional model are presented. Model parameter estimation with the Bayesian Markov chain Monte Carlo (MCMC) method is demonstrated. Then, the PISA 2012 computer-based mathematics data are analyzed to demonstrate the advantages of the proposed joint model. After this, a brief simulation study is conducted to evaluate the parameter recovery of the proposed model. Finally, conclusions and discussions are presented.

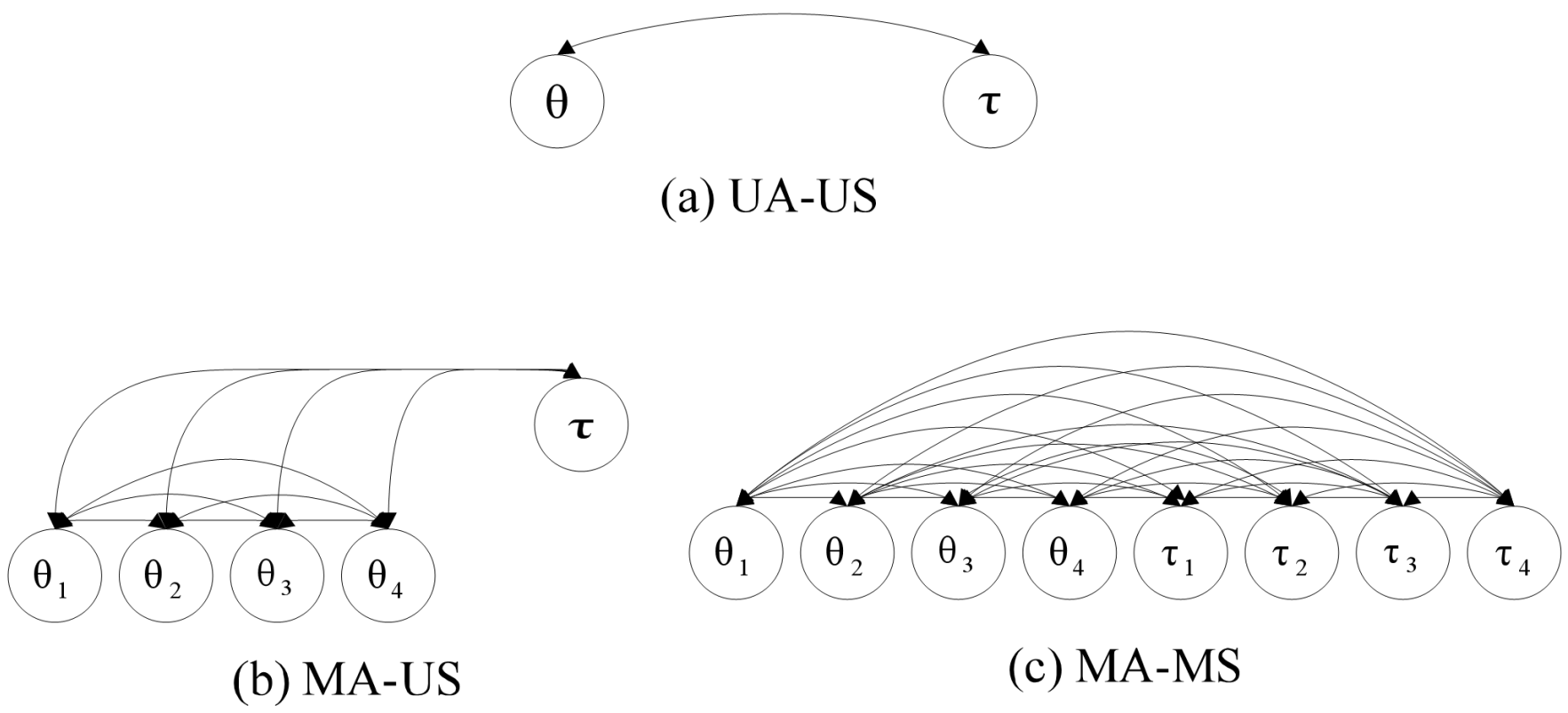

**Figure 1.** Three relationships between ability and speed in the hierarchical modeling framework. *Note*, U = unidimensional, M = multidimensional, A = ability, S = speed.

## Proposed Measurement Model for Response Time

### Overview of the Unidimensional Log-normal RT Model

Let $T_{ni}$ be the observed RT of person $n$ ($n$ = 1,..., $N$) to item $i$ ($i$ = 1,..., $I$). In the ULRTM, the logarithm function is used to transform the positively skewed distribution of RT to a symmetric shape and is assumed to be dominated by item $i$'s time-intensity parameter $\xi_i$ and person $n$'s latent speed parameter $\tau_n$ as follows:

$$\log T_{ni} = \xi_i - \tau_n + \varepsilon_{ni},\ \varepsilon_{ni} \sim N(0,\ \omega_i^{-2}), \tag{1}$$

or equivalently,

$$\log T_{ni} \sim N(\xi_i - \tau_n,\ \omega_i^{-2}). \tag{2}$$

where $\xi_i$ represents the time needed to complete item $i$; $\tau_n$ represents the working speed of person $n$ on a test and is assumed to be normally distributed with mean zero and variance $\sigma_\tau^2$; $\varepsilon_{ni}$ is the normally distributed residual error term, with mean zero and variance $\omega_i^{-2}$; $\omega_i$ is the reciprocal of the standard deviation of the error term, which can be treated as a time-kurtosis parameter.

A basic assumption of the ULRTM is that $\log T_{ni}$s are conditionally independent given the unidimensional $\tau_n$, which is known as local RT independence (Zhan, Liao et al., 2018). In other words, local RT independence is obtained when the relationship among items is fully characterized by the URT model. However, if the single latent speed is not sufficient to account for the relationship among RTs, an MRT model is needed.

**The Multidimensional Log-normal RT Model**

There are two types of multidimensional tests: between-item and within-item (Adams, Wilson, & Wang, 1997). In a between-item multidimensional test, each item measures a single dimension but different items may measure different dimensions so the multidimensionality occurs between items. In a within-item multidimensional test, an item may measure more than one dimension simultaneously. As the between-item multidimensionality can be seen as a special case of the within-item multidimensionality, to be general, we focus on the latter throughout this paper.

For an item that measures multiple dimensions simultaneously, the ULRTM can be extended to accommodate multiple latent speeds by replacing the scalar latent speed parameter, $\tau_n$, with a vector of speed parameter, $\boldsymbol{\tau}_n$. The resulting MLRTM is defined as:

$$\log T_{ni} = \xi_i - \mathbf{q}_i'\boldsymbol{\tau}_n + \varepsilon_{ni} = \xi_i - \sum_{k=1}^{K} q_{ik}\tau_{nk} + \varepsilon_{ni},\ \varepsilon_{ni} \sim N(0,\ \omega_i^{-2}), \quad (3)$$

or equivalently,

$$\log T_{ni} \sim N(\xi_i - \sum_{k=1}^{K} q_{ik}\tau_{nk},\ \omega_i^{-2}), \quad (4)$$

where $\tau_{nk}$ represents the latent speed of person $n$ in dimension $k$, and $\boldsymbol{\tau}_n = (\tau_{n1}, ..., \tau_{nk}, ..., \tau_{nK})'$ is the multidimensional latent speed vector following a multivariate normal distribution: $\boldsymbol{\tau}_n \sim N(\boldsymbol{\mu}_\tau, \boldsymbol{\Sigma}_\tau)$ with mean vector $\boldsymbol{\mu}_\tau$ and variance and covariance matrix $\boldsymbol{\Sigma}_\tau$, and $\boldsymbol{\mu}_\tau$ is set to $\mathbf{0}$-vector for identification; $q_{ik}$ is an element of the confirmatory Q-matrix (Tatsuoka, 1983) indicating whether dimension $k$ is required to answer item $i$ correctly; $q_{ik} = 1$ if the dimension is required, and 0 otherwise. Other parameters are the same as those in the ULRTM.

In the MLRTM, it is assumed that $\log T_{ni}$s are conditionally independent given $\boldsymbol{\tau}_n$. In addition, for a given item $i$, if $\sum_{k=1}^{K} \tau_{nk}$ is set at a constant, $m$, all $\boldsymbol{\tau}$-vectors that satisfy the expression $m = \sum_{k=1}^{K} \tau_{nk}$ yield the same RT. This feature suggests that a low speed on one dimension can be compensated by a high speed on another dimension, so the proposed MLRTM is a compensatory MLRTM, which is in line with the compensatory assumption of latent abilities in compensatory multidimensional item response theory (MIRT) models. Logically speaking, within each dimension, the latent speed and the latent ability should be matched with each other (e.g., Zhan, Liao et al., 2018). Thus, when the latent abilities are compensatory, it is reasonable to assume that the corresponding latent speeds to be compensatory as well. On the other hand, if the latent abilities are non-compensatory, the corresponding latent speeds should be non-compensatory as well. However, the development of non-compensatory MLRTM is beyond the scope of this study, and can be studied in the future.

## Multidimensional Hierarchical Modeling Framework

Since both response accuracy and RTs contain information about items and persons, it is advantageous to analysis them simultaneously. For example, one may be interested in the relationship between multidimensional latent ability and multidimensional latent speed of persons and the relationship between difficulty and time-intensity of items.

In the multidimensional hierarchical modeling framework, at the first level, an MIRT model can be used as the measurement model for response accuracy and an MRT model can be used as the measurement model for RTs, respectively; at the second level, all latent abilities and latent speeds are assumed to follow a multivariate normal distribution; meanwhile, the item RA parameters (e.g., item difficulty) and item RT parameters (e.g., item time-intensity) are assumed to follow a multivariate normal distribution. Given the “plug-and-play” nature of the multidimensional hierarchical modeling, various choices of MIRT models and MRT models can be adopted. For illustration purposes, in this study, the MLRTM is used as the measurement model for RTs, and according to the 2012 PISA mathematics assessment framework (OECD, 2013), the multidimensional Rasch model (MRM; Adams et al., 1997) was employed as the measurement model for response accuracy.

**The Multidimensional Rasch Model for RA**

Let $Y_{ni}$ be the observed response for person $n$ to item $i$. The MRM can be expressed as

$$\text{logit}(P(Y_{ni}=1)) = \mathbf{q}_i'\boldsymbol{\theta}_n + d_i = \sum_{k=1}^{K} q_{ik}\theta_{nk} + d_i\,, \tag{5}$$

where logit($x$) = log($x$/(1–$x$)); $P(Y_{ni} = 1)$ is the probability of a correct response by person $n$ to item $i$; $\theta_{nk}$ is the latent ability of person $n$ on dimension $k$, and $\boldsymbol{\theta}_n = (\theta_{n1}, ..., \theta_{nk}, ..., \theta_{nK})'$ is assumed to follow a multivariate normal distribution as follows: $\boldsymbol{\theta}_n \sim N(\boldsymbol{\mu}_\theta, \boldsymbol{\Sigma}_\theta)$, and setting $\boldsymbol{\mu}_\theta = \mathbf{0}$ for identification; $d_i$ is the intercept or easiness of item $i$; $q_{ik}$ is an element of the confirmatory Q-matrix indicating whether dimension $k$ is required to answer item $i$ correctly.

Further, if only one dimension is assumed to be required by all items, the MRM reduces to the unidmensional Rasch model (URM, Rasch, 1960),

$$\text{logit}(P(Y_{ni}=1)) = \theta_n + d_i\,, \tag{6}$$

where $\theta_n$ is the unidimensional latent ability of person $n$; $d_i$ is the intercept or easiness of item $i$. The URM can be identified by setting the mean of $\theta_n$ at zero.

**Multidimensional Hierarchical Modeling**

In multidimensional tests, three possible structures/combinations of latent speed and latent ability have been outlined in Figure 1, including (a) unidimensional ability and unidimensional speed (UA-US), in which a single latent ability is associated with a single latent speed; (b) multidimensional ability and unidimensional speed (MA-US), in which four (in this example) latent abilities are associated with a single latent speed; and (c) multidimensional ability and multidimensional speed (MA-MS), in which four latent abilities are associated with four latent speeds. Among the three structures, the MA-MS structure is the most general and contains the other two as special cases. Although the UA-US structure is commonly used in unidimensional tests, it still can be compulsively used in multidimensional tests by assuming all items required only a single dimension. The MA-US structure was proposed by Man et al. (2017) and Zhan, Jiao et al. (2018). The MA-MS structure is proposed in this study.

Different structures represent different combinations of measurement models. The UA-US represents a combination of the URM and the ULRTM, the MA-US represents a combination of the MRM and the ULRTM, and the MA-MS represents a combination of the MRM and the MLRTM. Besides the measurement models, the multivariate normal distribution was used to describe the relationship among the latent abilities and latent speeds:

$$\mathbf{\Omega}_n = \begin{pmatrix} \boldsymbol{\theta}_n \\ \boldsymbol{\tau}_n \end{pmatrix} \sim N_{K*}\left( \begin{pmatrix} \boldsymbol{\mu}_\theta \\ \boldsymbol{\mu}_\tau \end{pmatrix}, \mathbf{\Sigma}_{\text{person}} \right), \quad \mathbf{\Sigma}_{\text{person}} = \begin{pmatrix} \sigma^2_{\theta_1} & & & & & \\ \vdots & \ddots & & & & \\ \sigma_{\theta_1\theta_K} & \cdots & \sigma^2_{\theta_K} & & & \\ \sigma_{\theta_1\tau_1} & \cdots & \sigma_{\theta_K\tau_1} & \sigma^2_{\tau_1} & & \\ \vdots & \cdots & \vdots & \vdots & \ddots & \\ \sigma_{\theta_1\tau_K} & \cdots & \sigma_{\theta_K\tau_K} & \sigma_{\tau_1\tau_K} & \cdots & \sigma^2_{\tau_K} \end{pmatrix}, \tag{7}$$

where $\boldsymbol{\theta}_n$ is a vector of multiple latent abilities of person $n$; $\boldsymbol{\tau}_n$ is a vector of multiple latent speeds of person $n$; $\mathbf{\Sigma}_{\text{person}}$ is a variance and covariance matrix of person parameters; $K^*$ is the total number of latent abilities and latent speeds. Take the structures in Figure 1 as examples, a bivariate normal distribution ($K^* = 2$) can be employed for the UA-US; a fivefold-variate normal distribution ($K^* = 5$) can be employed for the MA-US; and an eightfold-variate normal distribution ($K^* = 8$) can be employed for the MA-MS.

For the item parameters, a bivariate normal distribution was used to describe the relationship between item difficulty and item time-intensity:

$$\boldsymbol{\Psi}_i = \begin{pmatrix} d_i \\ \xi_i \end{pmatrix} \sim N_2\left(\begin{pmatrix} \mu_d \\ \mu_\xi \end{pmatrix}, \boldsymbol{\Sigma}_{\text{item}}\right), \quad \boldsymbol{\Sigma}_{\text{item}} = \begin{pmatrix} \sigma_d^2 & \\ \sigma_{d\xi} & \sigma_\xi^2 \end{pmatrix}, \tag{8}$$

where $\mu_d$ and $\mu_\xi$ is the mean of item difficulty and item time-intensity, respectively; $\boldsymbol{\Sigma}_{\text{item}}$ is a variance and covariance matrix of item parameter. The residual error variance, $\omega_i^{-2}$, is assumed to be independently distributed (e.g., Zhan, Jiao et al., 2018), thus it is not included in $\boldsymbol{\Psi}_i$.

**Bayesian parameter estimation**

Model parameters in the MRM-MLRTM can be estimated via the full Bayesian approach with the Markov chain Monte Carlo (MCMC) method. In Bayesian estimation, prior distributions of model parameters and observed data likelihood produce a joint posterior distribution for the model parameters. In this study, the OpenBUGS (Spiegelhalter, Thomas, Best, & Lunn, 2014) was used to estimate parameters. OpenBUGS uses a default option of the Gibbs sampler (Gelfand & Smith, 1990), whose code for the MRM-MLRTM is provided in online supplemental materials (runnable source code are also provided).

Under the assumption of local independence, $Y_{ni}$ and $\log T_{ni}$ are independently distributed as

$$Y_{ni} \sim \text{Bernoulli}(P(Y_{ni} = 1)), \text{ and } \log T_{ni} \sim N(\xi_i - \textstyle\sum_{k=1}^{K} q_{ik}\tau_{nk},\ \omega_i^{-2}).$$

The priors of the person parameters are set as

$$\begin{pmatrix} \boldsymbol{\Theta}_n \\ \mathbf{T}_n \end{pmatrix} \sim N_{K^*}\left(\begin{pmatrix} \mathbf{0} \\ \mathbf{0} \end{pmatrix}, \boldsymbol{\Sigma}_{\text{person}}\right),$$

with a hyper prior

$$\boldsymbol{\Sigma}_{\text{person}} \sim \text{InvWishart}(\mathbf{R}_{\text{person}}, K^*),$$

where $\mathbf{R}_{\text{person}}$ is a $K^*$-dimensional identity matrix.

In addition, the priors of item parameters are set as

$$\begin{pmatrix} d_i \\ \xi_i \end{pmatrix} \sim N_2\left(\begin{pmatrix} \mu_d \\ \mu_\xi \end{pmatrix}, \mathbf{\Sigma}_{\text{item}}\right), \quad \omega_i^{-2} \sim \text{InvGamma}(1,\ 1),$$

Furthermore, the hyper priors are specified as:

$$\mu_d \sim \text{Normal}(0,\ 2), \quad \mu_\xi \sim \text{Normal}(4.3,\ 2), \quad \mathbf{\Sigma}_{\text{item}} \sim \text{InvWishart}(\mathbf{R}_{\text{item}}, 2),$$

where $\mathbf{R}_{\text{item}}$ is a two-dimensional identity matrix. Finally, the posterior mean is treated as the estimate for model parameters.

# Real Data Analysis

## Data Description

A real data analysis is conducted using the PISA 2012 computer-based mathematics data to explore whether the MRM-MLRTM fit the data better than the URM-ULRTM (e.g., van der Linden, 2007) and MRM-ULRTM (e.g., Man et al., 2017) when the test structure is multidimensional. It is also an example to illustrate the use of the proposed model.

This dataset was used by Zhan, Jiao et al. (2018). There are $N = 1581$ respondents and $I = 10$ items. The logarithm of RTs have been computed before analysis, and all zero RTs were treated as missing data. Four dimensions that belong to the mathematical content knowledge were chosen in this study, namely, change and relationships ($\theta_1$), quantity ($\theta_2$), space and shape ($\theta_3$), and uncertainty and data ($\theta_4$). The Q-matrix is shown in Table 1. Due to the use of 10 items to measure four dimensions and the incompleteness and unidentifiability of the Q-matrix (Chiu, 2013; Xu & Zhang, 2016), it is expected that the estimation of model parameters may contain relatively high measurement errors. Nevertheless, such a test structure was retained because, the real data analysis, though not perfect, can still provide some information about the nature of the real data structure. In addition, Bayesian estimation works even when the model is not identified, and adherence to identifiability has been deemed to be largely superfluous (Gustafson et al., 2005; Muthén & Asparouhov, 2012). More details can be found in Zhan, Jiao et al. (2018).

Note that we also analyzed the RT data by using the MLRTM alone, the results indicated that the MLRTM fit the RT data better than the ULRTM. More details can be found in online supplemental materials.

**Table 1.** Q matrix for PISA 2012 released computer-based mathematics items.

| Items | $\theta_1$ | $\theta_2$ | $\theta_3$ | $\theta_4$ |
|---|---|---|---|---|
| CM015Q01 | | 1 | | |
| CM015Q02D | 1 | | | |
| CM015Q03D | 1 | | | |
| CM020Q01 | | | 1 | |
| CM020Q02 | | | 1 | |
| CM020Q03 | | | 1 | |
| CM020Q04 | | | 1 | |
| CM038Q03T | | | | 1 |
| CM038Q05 | | | | 1 |
| CM038Q06 | | | | 1 |

*Note*, blank means zero.

**Analysis and Model Selection**

The three joint models in Figure 1, namely, the URM-ULRTM, the MRM-ULRTM, and the MRM-MLRTM were fit to the data. For each model, two Markov chains with random starting points were used, and each chain ran 10,000 iterations with the first 5,000 iterations in each chain as burn-in. Finally, the remaining 10,000 iterations were used for the model parameter inferences. The potential scale reduction factor (PSRF; Brooks & Gelman, 1998) was computed to assess the convergence of each parameter. A PSRF with values smaller than 1.1 or 1.2 indicates convergence (Brooks & Gelman, 1998; Zhan, Liao et al., 2018). Our studies indicated that the PSRF was smaller than 1.05 for all parameters, suggesting good convergence.

Posterior predictive model checking (PPMC; Gelman, Carlin, Stern, Dunson, Vehtari, &

Rubin, 2014) was used to evaluate model-data fit. A posterior predictive probability (*ppp*) value near 0.5 indicates that there are no systematic differences between the realized and predictive values, and thus an adequate fit of the model. In PPMC, the sum of the squared Pearson residuals for person *n* and item *i* (Yan, Mislevy, & Almond, 2003) was used as a discrepancy measure to evaluate the overall fit of the RA model, which is written as

$$D(Y_{ni};\boldsymbol{\alpha}_n,\beta_i,\delta_i)=\sum_{n=1}^{N}\sum_{i=1}^{I}\left(\frac{Y_{ni}-P(Y_{ni}=1)}{\sqrt{P(Y_{ni}=1)(1-P(Y_{ni}=1))}}\right)^2,$$

where $P(Y_{ni} = 1)$ has the same definition as that in Equation (5). On the other hand, the sum of the standardized error function of $\log T_{ni}$ for person *n* and item *i* was employed as a discrepancy measure:

$$D(\log\mathbf{T};\boldsymbol{\upsilon})=D(\log T_{ni};\xi_i,\boldsymbol{\tau}_n,\omega_i)=\sum_{n=1}^{N}\sum_{i=1}^{I}\left(\omega_i(\log T_{ni}-(\xi_i-\sum_{k=1}^{K}q_{ik}\tau_{nk}))\right)^2.$$

Additionally, the Akaike Information Criterion (AIC; Akaike, 1974), the Bayesian Information Criterion (BIC; Schwar, 1978), and the deviance information criterion (DIC; Spiegelhalter, Best, Carlin, & van der Linde, 2002) were computed for model selection. Specifically, $DIC=\overline{D}+p_e=\overline{D}+\mathrm{var}(D)/2$, namely, the effective number of parameters ($p_e$) was computed by $p_e=\mathrm{var}(D)/2$ (Gelman, Carlin, Stern, & Rubin, 2003), where $D$ is the deviance, and $\overline{D}$ is the posterior mean of deviance (i.e., –2 log likelihood). Note that in Bayesian analysis, the AIC and BIC can be defined as $AIC=\overline{D}+p$ and $BIC=\overline{D}+(\log N-1)p$ (Congdon, 2003), where $p$ is the number of estimated parameters. A smaller value of indicates a better model-data fit.

**Results**

The AIC, BIC, and DIC all identified the MRM-MLRTM as the best-fitting model and the URM-ULRTM as the worst-fitting model, as shown in Table 2. In the MRM-MLRTM, the *ppp* values of the RA model and the RT model were 0.789 and 0.616, respectively, which indicated an adequate model-data fit. Thus, it may be more appropriate to simultaneously consider the multidimensionality of the latent speeds and latent abilities. The MRM-MLRTM was used for further illustration.

**Table 2.** Models Fit for the PISA 2012 Computer-based Mathematics Data.

| Analysis Model | | AIC | BIC | DIC | *ppp*_RA | *ppp*_RT |
|---|---|---|---|---|---|---|
| For Ability | For Speed | | | | | |
| MRM | MLRTM | **33434** | **33810** | **38456** | 0.789 | 0.616 |
| MRM | ULRTM | 35188 | 35451 | 39335 | 0.566 | 0.568 |
| URM | ULRTM | 38130 | 38329 | 40552 | 0.813 | 0.559 |

*Note*, MRM = multidimensional Rasch model; MLRTM = multidimensional log-normal response time model; ULRTM = unidimensional log-normal response time model; AIC = Akaike information criterion; BIC = Bayesian information criterion; DIC = deviance information criterion; ppp = posterior predictive probability value; RA = item response accuracy; RT = item response times.

To explore the relationship among multiple the latent abilities and latent speeds, Table 3 presents the estimated person variance and covariance matrix. Moderate to high positive correlations (0.70 ~ 0.92) were found among the multiple abilities, and lower correlations (0.56 ~ 0.85) were found among the multiple speeds. However, low to moderate negative correlations (–0.60 ~ –0.01) were found between the multidimensional abilities and speeds. Although these results are not consistent with common sense that more capable respondents tended to work faster, they are consistent with previous study findings (Klein Entink, Fox et al., 2009; van der Linden & Fox, 2015; Zhan, Jiao et al., 2018). As a low-stakes test, the PISA has more significant implications for countries or areas than individual respondents. A reasonable explanation could be that low ability respondents lacked motivation in taking the low-stakes test (Wise & Kong, 2005), which led to shorter RTs. Note that the variance of the first latent ability was quite large, indicating all respondents differ greatly in the first dimension.

**Table 3.** Estimated Variance and Covariance Matrix for the Multidimensional Abilities and Speeds.

| | $\theta_1$ | $\theta_2$ | $\theta_3$ | $\theta_4$ | $\tau_1$ | $\tau_2$ | $\tau_3$ | $\tau_4$ |
|---|---|---|---|---|---|---|---|---|
| $\theta_1$ | 11.996 (2.480) | 0.916 | 0.774 | 0.822 | –0.629 | –0.008 | –0.260 | –0.348 |
| $\theta_2$ | 5.805 (1.118) | 3.345 (0.943) | 0.696 | 0.721 | –0.558 | –0.104 | –0.188 | –0.264 |
| $\theta_3$ | 3.790 (0.507) | 1.799 (0.245) | 1.999 (0.216) | 0.842 | –0.467 | –0.060 | –0.299 | –0.417 |
| $\theta_4$ | 2.886 (0.421) | 1.337 (0.212) | 1.207 (0.123) | 1.027 (0.125) | –0.600 | –0.075 | –0.392 | –0.569 |
| $\tau_1$ | –1.209 (0.187) | –0.567 (0.080) | –0.367 (0.041) | –0.338 (0.035) | 0.308 (0.017) | 0.563 | 0.739 | 0.753 |
| $\tau_2$ | –0.014 (0.086) | –0.099 (0.049) | –0.044 (0.031) | –0.040 (0.027) | 0.163 (0.012) | 0.272 (0.027) | 0.651 | 0.605 |
| $\tau_3$ | –0.404 (0.093) | –0.154 (0.037) | –0.190 (0.028) | –0.179 (0.025) | 0.184 (0.011) | 0.152 (0.010) | 0.202 (0.010) | 0.850 |
| $\tau_4$ | –0.654 (0.125) | –0.262 (0.051) | –0.319 (0.038) | –0.312 (0.034) | 0.227 (0.013) | 0.171 (0.012) | 0.207 (0.011) | 0.294 (0.014) |

*Note*, standard error in parentheses.

Figure 2 presents four scatter diagrams to further depict the relationship between the multidimensional construct and the corresponding multidimensional speed. Obviously, different relationships existed in different dimensions. A relatively fuzzy relationship in the second dimension might be caused the unstable estimation based on a single item. According to the scatter diagrams, some respondents had low ability but high speed, found in the fourth quadrant, and they might demonstrate aberrant response behavior such as rapid-guessing (Fox & Marianti, 2017; Wang & Xu, 2015).

Table 4 presents the estimates of the item parameters. The estimated mean item difficulty and mean time-intensity were –1.19 (SE = 0.54) and 4.29 (SE = 0.15), respectively. The estimates of the time-intensity and time-kurtosis were quite similar to those shown in Table S3 in online supplemental materials, which were estimated from the MLRTM alone. To further explore the relationship between the item intercept and time-intensity parameters, we present the estimated item variance and covariance matrix in Table 5. The correlation was –0.43, which implied that the items with a lower intercept (i.e., more difficult) might have higher time-intensity. This result was consistent with that in the literature in that the more difficult items often need more time to solve (Fox & Marianti, 2016; Meng et al., 2015; van der Linden, 2006; 2007).

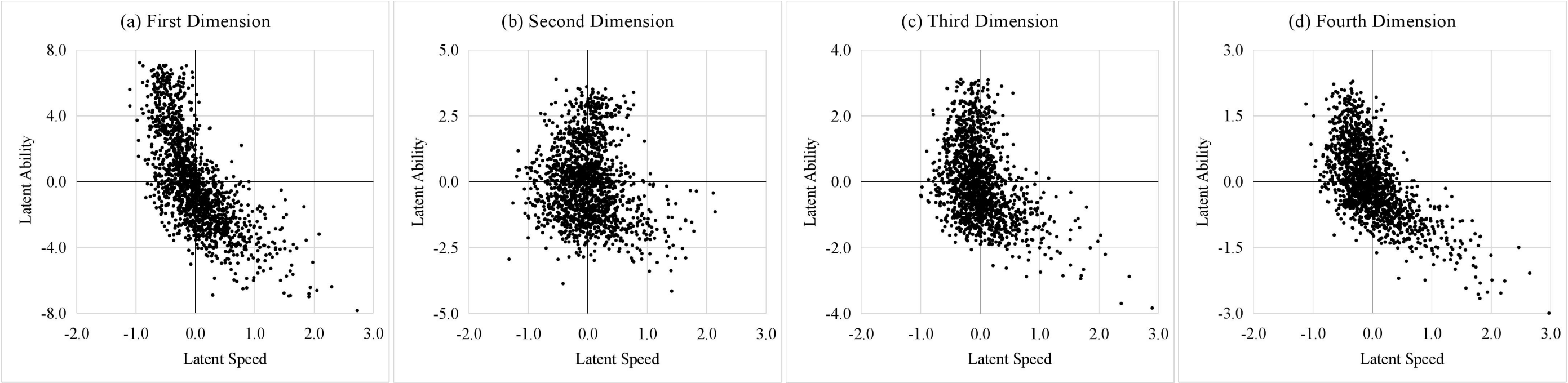

**Figure 2**. Relationship between the multidimensional construct and the corresponding multidimensional speed.

**Table 4.** Estimated Item Parameters for the Released 2012 PISA Computer-based Mathematics Items.

| Item | *d* | | ξ | | ω | |
|---|---|---|---|---|---|---|
| | Mean | SE | Mean | SE | Mean | SE |
| 1 | 0.567 | 0.095 | 4.224 | 0.016 | 2.865 | 0.026 |
| 2 | –4.548 | 0.463 | 4.472 | 0.020 | 1.865 | 0.015 |
| 3 | –3.415 | 0.358 | 4.631 | 0.019 | 1.995 | 0.015 |
| 4 | –2.384 | 0.114 | 4.779 | 0.015 | 2.524 | 0.008 |
| 5 | –0.129 | 0.069 | 3.861 | 0.017 | 1.913 | 0.012 |
| 6 | –1.441 | 0.087 | 4.258 | 0.016 | 2.190 | 0.010 |
| 7 | –0.419 | 0.072 | 3.739 | 0.017 | 2.097 | 0.010 |
| 8 | 0.749 | 0.066 | 4.190 | 0.017 | 2.562 | 0.011 |
| 9 | –1.380 | 0.077 | 4.523 | 0.018 | 2.085 | 0.012 |
| 10 | –1.438 | 0.077 | 4.380 | 0.021 | 1.689 | 0.016 |

*Note*, Mean = posterior mean; SE = standard deviation of posterior distribution.

**Table 5.** Estimated Variance and Covariance Matrix of Item Parameters for the Released 2012 PISA Computer-based Mathematics Items.

| | *d* | ξ |
|---|---|---|
| *d* | 3.376 (2.018) | –0.433 |
| ξ | –0.391 (0.370) | 0.242 (0.135) |

*Note*, covariance in lower triangular and correlation coefficient in upper triangular.

## A Brief Simulation Study for Parameter Recovery

### Design and Data Generation

A real data study has been provided to demonstrate the applications of the proposed joint multidimensional model. Further, to assess the parameter recovery of the proposed model, a brief simulation study was conducted. For simplicity, only the MRM-MLRTM was assessed, because other two joint hierarchical models can be seen as its special cases.

Four dimensions were measured by 30 items, which means there are four latent abilities and four corresponding latent speeds. Q matrix was presented in Figure 3. For item parameters, $d$ and $\xi$ parameters generated from a bivariate normal distribution with mean vector (0, 4) and covariance matrix of [1, –0.2; –0.2, 0.25], in such setting, $\rho_{d\xi} = -0.4$. $\omega$ parameters were set to 2, which were also similar to the estimates in the real data analysis. 1,000 respondents were simulated. ($\mathbf{\Theta}$, $\mathbf{T}$) generated from an eightfold-variate normal distribution with mean vector ($\mathbf{0}$, $\mathbf{0}$) and covariance matrix of

$$\begin{pmatrix} 1 & & & & & & & \\ 0.8 & 1 & & & & & & \\ 0.8 & 0.8 & 1 & & & & & \\ 0.8 & 0.8 & 0.8 & 1 & & & & \\ -0.4 & -0.4 & -0.4 & -0.4 & 1 & & & \\ -0.4 & -0.4 & -0.4 & -0.4 & 0.6 & 1 & & \\ -0.4 & -0.4 & -0.4 & -0.4 & 0.6 & 0.6 & 1 & \\ -0.4 & -0.4 & -0.4 & -0.4 & 0.6 & 0.6 & 0.6 & 1 \end{pmatrix}.$$

In such settings, $\rho_{\theta\theta'} = 0.8$, $\rho_{\tau\tau'} = 0.6$, and $\rho_{\theta\tau} = -0.4$. 30 datasets were generated.

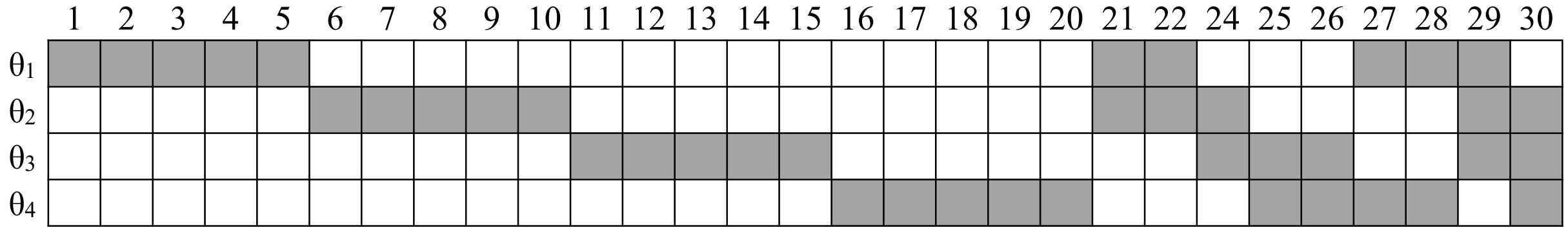

**Figure 3.** *K*-by-*I* Q' matrix in the simulation study.

**Analysis**

The MRM-MLRTM was fitted to each of the 30 replications. In each replication, the numbers of chains, burn-in iterations and post-burn-in iterations were the same as those set in the empirical study. It appeared that convergence was well achieved. To evaluate parameter recovery, the bias and the root mean square error (RMSE) were computed as: $\text{bias}(\hat{\upsilon}) = \sum_{r=1}^{R} \frac{\hat{\upsilon}_r - \upsilon}{R}$ and $\text{RMSE}(\hat{\upsilon}) = \sqrt{\sum_{r=1}^{R} \frac{(\hat{\upsilon}_r - \upsilon)^2}{R}}$, where $\hat{\upsilon}$ and $\upsilon$ are the estimated and true values of model parameters, respectively; $R$ is the total number of replications. The correlation between estimated and true values (Cor) was also computed.

**Results**

Table 6 presents the bias, RMSE, and the Cor of item and person parameters. Across the board, all model parameters are well recovered. For item parameters, the recovery of time-intensity is the best, then is time-kurtosis, and the worst is item intercept. For person parameters, the recovery of latent speeds is better than that of latent abilities. The recovery of variance and covariance was of more interest in this study; estimated bias and RMSE are given in Table 7 for $\mathbf{\Sigma}_{\text{item}}$ and Tables 8 and 9 for $\mathbf{\Sigma}_{\text{person}}$, respectively. In general, all of them are well recovered. The recovery of covariances is better than that of variances, and the recovery of time-related parameters (e.g., item intensity, covariance of item difficulty and time-intensity, latent speeds, and covariance of latent ability and latent speed) is better than that of time-unrelated parameters (e.g., item intercept and latent abilities). Overall, model parameters of the MRM-MLRTM can be recovered very well via the proposed full Bayesian MCMC estimation algorithm.

**Table 6.** Recovery of Person Parameters in the Simulation Study.

| Parameter | Mean Bias | Mean RMSE | Cor |
|---|---|---|---|
| $d$ | 0.002 | 0.078 | 0.997 |
| $\xi$ | 0.002 | 0.018 | 0.999 |
| $\omega$ | –0.011 | 0.049 | NA |
| $\theta_1$ | –0.001 | 0.468 | 0.886 |
| $\theta_2$ | –0.001 | 0.458 | 0.888 |
| $\theta_3$ | –0.001 | 0.460 | 0.886 |
| $\theta_4$ | –0.001 | 0.467 | 0.883 |
| $\tau_1$ | 0.001 | 0.161 | 0.987 |
| $\tau_2$ | 0.002 | 0.161 | 0.987 |
| $\tau_3$ | 0.001 | 0.161 | 0.987 |
| $\tau_4$ | 0.001 | 0.161 | 0.987 |

*Note*, Mean Bias = mean bias across all respondents; Mean RMSE = mean RMSE acorss all respondents; Cor = correlation between estimated and true values; Cor of ω is NA because of the variance of true ω is zero.

**Table 7.** Recovery of Item Mean Vector and Item Variance and Covariance Matrix.

| Parameter | Bias | RMSE |
|---|---|---|
| $\sigma_d^2$ | 0.070 | 0.080 |
| Cov($d$, $\xi$) | –0.005 | 0.011 |
| $\sigma_\xi^2$ | 0.043 | 0.043 |
| $\mu_d$ | 0.003 | 0.015 |
| $\mu_\xi$ | 0.002 | 0.009 |

**Table 8.** Bias of the Variance and Covariance Matrix of Person Parameters.

| $\Sigma_{person}$ | $\theta_1$ | $\theta_2$ | $\theta_3$ | $\theta_4$ | $\tau_1$ | $\tau_2$ | $\tau_3$ | $\tau_4$ |
|---|---|---|---|---|---|---|---|---|
| $\theta_1$ | –0.043 | | | | | | | |
| $\theta_2$ | –0.002 | –0.007 | | | | | | |
| $\theta_3$ | –0.016 | 0.000 | –0.046 | | | | | |
| $\theta_4$ | –0.011 | 0.006 | 0.003 | –0.026 | | | | |
| $\tau_1$ | 0.002 | –0.001 | 0.003 | –0.005 | 0.005 | | | |
| $\tau_2$ | 0.005 | 0.003 | 0.005 | 0.005 | 0.000 | –0.003 | | |
| $\tau_3$ | 0.006 | –0.002 | 0.006 | –0.001 | 0.004 | 0.001 | 0.002 | |
| $\tau_4$ | 0.005 | –0.002 | –0.003 | –0.003 | 0.000 | –0.001 | 0.002 | 0.000 |

**Table 9.** RMSE of the variance and covariance matrix of person parameters.

| $\Sigma_{person}$ | $\theta_1$ | $\theta_2$ | $\theta_3$ | $\theta_4$ | $\tau_1$ | $\tau_2$ | $\tau_3$ | $\tau_4$ |
|---|---|---|---|---|---|---|---|---|
| $\theta_1$ | 0.083 | | | | | | | |
| $\theta_2$ | 0.048 | 0.075 | | | | | | |
| $\theta_3$ | 0.052 | 0.039 | 0.081 | | | | | |
| $\theta_4$ | 0.042 | 0.044 | 0.051 | 0.074 | | | | |
| $\tau_1$ | 0.031 | 0.023 | 0.029 | 0.031 | 0.013 | | | |
| $\tau_2$ | 0.026 | 0.031 | 0.030 | 0.031 | 0.008 | 0.012 | | |
| $\tau_3$ | 0.027 | 0.025 | 0.031 | 0.034 | 0.010 | 0.007 | 0.014 | |
| $\tau_4$ | 0.026 | 0.023 | 0.031 | 0.030 | 0.009 | 0.007 | 0.007 | 0.010 |

## Conclusion and discussion

To capture the multidimensionality of latent speed, this study proposed a multidimensional log-normal RT model and a multidimensional hierarchical modeling framework. The PISA 2012 computer-based mathematics data were analyzed to illustrate the implications and applications of the proposed models. The results indicating that it is appropriate to consider the multidimensionality of latent speed and the multidimensionality of latent ability, simultaneously, in multidimensional tests when RTs were collected. A brief simulation study was used as well to further evaluate model parameter recovery. The results indicated that model parameters could be well recovered using the Bayesian MCMC approach.

The work presented in this article is only a first attempt to deal with the multidimensionality of latent speed. Despite promising results, further exploration is encouraged. First, the proposed RT model is a multidimensional extension of the classical log-normal RT model (van der Linden, 2006), multidimensional extensions of other possible RT models (Fox & Marianti, 2016; Klein Entink, van der Linden, et al., 2009; Wang, et al., 2013; Wang & Xu, 2015) could be explored and compared in the future. Second, in the proposed MLRTM, the latent speeds are assumed to be compensatory. As non-compensatory multidimensional models for response accuracy become popular in recent decades (DeMars, 2016; Embretson, 1984; 2015; Templin & Henson, 2006;

Wang & Nydick, 2015; Jiao, Lissitz, & Zhan, 2017), it is important to develop corresponding non-compensatory MRT models in the future. Third, in the proposed multidimensional hierarchical modeling approach, a multivariate normal distribution was used to describe the relationships among multidimensional latent speed and multidimensional latent ability. So, the number of total dimensions is twice as many as the number of dimensions that are measured by the test. For example, in above application example, there were eight dimensions, which may pose a challenge on parameter estimation. If the multidimensional latent ability and the multidimensional latent speed can each have a second-order (or bi-factor) structure, not only the parameter estimation challenge can be largely reduced but also the structures of latent ability and latent speed can be posited and tested. Fourth, applications of the proposed RT model, like detecting aberrant responses (e.g., rapid-guessing and cheating) in multidimensional tests, need further investigation. Fifth, analyzing students' growth is an important topic in educational and psychological research (e.g., von Davier, Xu, & Carstensen, 2011; Zhan, Jiao, & Liao, 2017). How to employ the proposed multidimensional hierarchical modeling approach into longitudinal studies is also an interesting topic (e.g., Wang, Zhang, Douglas, & Culpepper, 2018).